\documentclass[preprint,12pt]{elsarticle}

\usepackage{graphicx}
\usepackage{subfigure}
\usepackage{setspace}
\usepackage{amssymb}
\usepackage{amsmath}
\usepackage{bm}

\usepackage{array,multirow}
\usepackage{float}
\usepackage[%
     colorlinks=true,
     urlcolor=blue,
     linkcolor=blue,
     citecolor=blue
]{hyperref}

\usepackage{lineno}
\usepackage{xcolor}
\usepackage[flushleft]{threeparttable}
\usepackage{comment}
\usepackage{enumitem}
\usepackage{xcolor}
\usepackage{soul,color}

\journal{Journal Name}

\begin{document}

\begin{frontmatter}


\title{Mechanism of keyhole pore formation in metal additive manufacturing}



\author[1]{Lu Wang}
\author[1]{Yanming Zhang}
\author[1]{Hou Yi Chia}
\author[1,*]{Wentao Yan}

\address[1]{Department of Mechanical Engineering, National University of Singapore, 117575, Singapore}
\address[*]{Corresponding author: mpeyanw@nus.edu.sg}

\begin{abstract}
Metal additive manufacturing has gained extensive attention from research institutes and companies to fabricate intricate parts and functionally graded materials.
However, the porosity of the as-built part deteriorates the mechanical property and even hinders the further application of metal additive manufacturing.
Particularly, the mechanisms of keyhole pores associated with the keyhole fluctuation are not fully understood.
To reveal the mechanisms of the keyhole pores formation, we adopt a multiphysics thermal-fluid flow model incorporating heat transfer, liquid flow, metal evaporation, Marangoni effect, and Darcy's law to simulate the keyhole pore formation process, and the results are validated with the in-situ X-ray images.
The simulation results present the instant bubble formation due to the keyhole instability and motion of the instant bubble when it pins on the solidification front.
Moreover, the unevenly distributed recoil pressure on the keyhole surface is an important factor for keyhole collapse and pentration.
Furthermore, comparing the keyhole pore formation under different laser scanning speeds shows that the keyhole pore is sensitive to the manufacturing parameters.
The keyhole fluctuation features and energy absorptivity variation on the rear keyhole wall could be metrics to evaluate the likelihood of the keyhole pore formation.
Additionally, the simulation under a low ambient pressure shows the feaibility of improving the keyhole stability to reduce and even avoid the formation of keyhole pores.

\end{abstract}


\end{frontmatter}


\section{Introduction}
\label{S:1}
Metal additive manufacturing (AM) is well known for its ability to fabricate complex-shaped parts without special tooling \cite{yang2012compressive} and functionally graded parts \cite{yan2016multi}, shorten the development cycle of products \cite{wei2021mechanistic}, and save the cost of the material \cite{debroy2018additive}.
However, the manufacturing defects \cite{debroy2018additive,snow2020review} are deleterious to the mechanical property.
The porosity of the as-built part, one kind of defects, decreases the ultimate strength directly and is also a  fatal flaw to the fatigue and fracture strength of the part \cite{beretta2017comparison,masuo2018influence,yang2018porosity}.
The presence of such defects does not meet the standards required in industry and thus prevents the adoption of AM technology in these industries.
Therefore, massive researches have been conducted to understand the pore formation mechanisms during AM and  control the as-built part porosity \cite{yang2018porosity,aboulkhair2014reducing,weingarten2015formation,king2014observation,leung2019effect,yang2020laser}.
Among various pore defects, the porosity under the keyhole mode melting \cite{king2014observation,cunningham2017synchrotron} is a ubiquitous defect in both laser welding \cite{fabbro2010melt} and laser powder bed fusion (L-PBF) and has attracted broad attention \cite{aboulkhair2014reducing,forien2020detecting,hojjatzadeh2020direct}.

Some progress has been made on the study of the keyhole pore features, keyhole collapse mechanisms, and keyhole pore formation process.
Ex-situ experiments \cite{debroy2018additive,snow2020review,king2014observation} show that the keyhole pores  are usually spherical and concentrate at the molten pool bottom.
However, these experiments have not observed the keyhole pore formation directly, and cannot present quantitative explanations for the keyhole pore formation mechanisms.
Recently, the in-situ X-ray imaging of keyhole dynamics \cite{hojjatzadeh2020direct,martin2019dynamics,zhao2020critical} were presented, and three types of keyhole pores were identified: 
(i) the instant bubble by a ledge on the rear keyhole  wall, 
(ii) the keyhole pore at the end of the track,
and (iii) the keyhole pore by the keyhole fluctuation. 
The first type of bubble is eliminated almost immediately after its formation and is insignificant for the pore formation mechanism.
The pores at the end of the track \cite{martin2019dynamics} are not only related to the keyhole dynamics but also determined by the laser scanning path, while these pores are usually reduced or eliminated by contour scanning and post-process polishing.
Thus, the keyhole pore by the keyhole fluctuation is the most significant, which is the focus of this study.

Besides the experimental approaches, numerical simulation of keyhole dynamics \cite{martin2019dynamics,khairallah2016laser,khairallah2020controlling,zhao2019bulk,wang2020evaporation} is a complementary, cost-saving and efficient approach to understand the formation mechanisms of the keyhole pores.
\citet{martin2019dynamics} simulated the keyhole pore at the end of the track and developed a strategy to reduce these pores.
The simulation by \citet{lin2017numerical} indicated that adjusting laser incident angle can reduce the number of keyhole pores during laser welding.
\citet{bayat2019keyhole} investigated the relationship between the keyhole pores and input power, and validated the simulation results with the experiment results.
The simulation by \citet{tang2018numerical} indicated that the spherical pores at the bottom of the molten pool formed as the energy density increased during AM.
The simulation by \citet{tan2021effects} showed the keyhole pore size decreased as the ambient pressure decreased.

Although the in-situ experiments and previous simulations have provided empirical observations on the keyhole pore formation, the basic principles like the recoil pressure distribution on the keyhole surface, the relationship between energy distribution and keyhole fluctuation, and molten pool flow during the keyhole pore formation remain elusive.
In this study, a multi-physics thermal-fluid flow model \cite{wang2020evaporation} incorporating heat transfer, molten pool flow, Marangoni effect, recoil pressure by metal evaporation, Darcy's law, and laser ray-tracing is adopted to simulate the keyhole fluctuation and keyhole pore formation process.
Firstly, we validate the simulation results of instant bubble formation and pinning on the solidification front against the X-ray imaging results.
Next, the varying trend of the keyhole depth fluctuation, absorbed energy distribution, keyhole pore size, molten pool flow, and forces on the keyhole with the increase of laser scanning speed are analyzed to explain the mechanisms and influence factors of the keyhole pore formation.
Furthermore, we explore the approach to reduce and even eliminate keyhole pores by simulating the molten pool flow under near-vacuum ambient pressure.
\section{Multiphysics thermal-fluid flow model}
The multiphysics thermal-fluid flow is based on our previous work \cite{wang2020evaporation,yan2017multi}. In the model, the liquid phase is assumed to be incompressible Newtonian fluid with laminar flow. 
The mass conservation equation is given as follows:
\begin{equation}
	\label{eq:continuity}
	\nabla \cdot \left ( \rho \boldsymbol{v}  \right )  = 0 
\end{equation}
\noindent where $\rho$ and $\boldsymbol{v}$  are the mass density and velocity vector.

The momentum conservation equation is given as follows:
\begin{equation}
	\label{eq:momentum-NS}
	\rho\frac{\partial \boldsymbol{v} }{\partial t} +\rho \nabla\cdot(\boldsymbol{v}\otimes\boldsymbol{v})
	= - \nabla p + \mu \nabla ^2 \boldsymbol{v} + \boldsymbol{f}_b - \rho D \boldsymbol{v}
\end{equation}      
\noindent where  $\mu$ and $p$ denote the dynamic viscosity and pressure. 
The buoyancy force $\boldsymbol{f}_b$ is accounted for using the Boussinesq approximation
\begin{equation}
    \label{eq:Fb}
    \boldsymbol{f}_b=\rho \boldsymbol{g} \alpha_v (T-T_{ref})
\end{equation}
where $\boldsymbol{g}$, $\alpha_v$, and $T_{ref}$ denote the gravitational acceleration vector, the thermal expansion coefficient, and reference temperature ( liquidus temperature $T_l$ for the current simulation). 
$D$ is the Darcy drag force coefficient, which is calculated by the Blake-Kozeny model \cite{poirier1987permeability,amador2016strategies}
\begin{equation}
    \label{eq:darcy}
    D = \frac{180\mu}{\rho \lambda_1^2} \frac{F_s^2}{(1-F_s)^3}
\end{equation}
where $\lambda_1$ is the characteristic length of mushy zone, taken as the primary dendrite arm spacing (about $5\;\rm{\mu m}$ for SLM \cite{wang2018additively}), and $F_s$ is the solid fraction. 

The energy conservation equation is given as follows:
\begin{equation}
	\label{eq:energy}
	\rho\frac{\partial I }{\partial t}+ \rho\nabla\cdot ( \boldsymbol{v} I)
	=\nabla\cdot (k \nabla T) +q
\end{equation}
\noindent where $k$ is the thermal conductivity, and $T$ is the temperature. $I=\int C_p \mathrm{d} T+(1-F_s)L_m$ is the specific internal energy, where $C_p$ and $L_m$ are the specific heat and specific latent heat of melting.  
$q$ is the power absorbed by the material, which is incorporated by using the ray-tracing method \cite{ahn2013three} to track the multireflections of laser and calculated with the Fresnel equation. In the current model, the reflection is assumed to be specular reflection.
The material for simulation in the current work is Ti-6Al-4V, but titanium's \cite{johnson1972optical} complex refractive index is used instead due to the lack of reliable data of Ti-6Al-4V.

The free surface of the molten pool is captured using the volume-of-fluid (VoF) method \cite{hirt1981volume}. 
\begin{equation}
	\label{eq:vof}
	\frac{\partial F }{\partial t}+ \nabla \cdot (F \boldsymbol{v})=0
\end{equation}
where $ F $ is the volume fraction. 

On the free surface, the normal ($p_\mathrm{n}$) and tangent forces ($\tau_\mathrm{t}$)  incorporate the surface tension, recoil pressure and Marangoni effect, given as 
\begin{equation}
\label{eq:bc_force}
\left\{
\begin{aligned}
    p_\mathrm{n} & = \sigma(T) \kappa +P_{rec}(T) \\
    \tau_\mathrm{t} & = \sigma_s^T \left[ \nabla T - \boldsymbol{n} (\nabla T \cdot \boldsymbol{n})\right]
\end{aligned}
\right.
\end{equation}
\noindent where $ \kappa $ is the curvature of the free surface, and $\boldsymbol{n}$ is the normal vector of the free surface. $ \sigma(T)=\sigma_0-\sigma_s^T(T-T_l ) $ and $ P_{rec}(T) $ are the temperature-dependent surface tension coefficient and  recoil pressure \cite{wang2020evaporation}. $ \sigma_0 $ and $ \sigma_s^T $ are  surface tension coefficient at the reference temperature $T_l$ (liquidus temperature in the current simulations) and its temperature sensitivity.

For the thermal boundary condition \cite{bayat2019keyhole}, it consists of heat convection , heat radiation, and heat loss of evaporation
\begin{equation}
\label{eq:bc_heat}
-k\nabla T\cdot \boldsymbol{n} = h(T-T_{env}) + \epsilon \sigma_s (T^4 - T^4_{env}) + \dot{q}_{evp}
\end{equation}
where $\sigma_s$ and $T_{env}$ are the Stefan-Boltzmann constant ($5.6704\times10^{-8}\;\rm{W/(m^2\cdot K^4)}$) and the ambient temperature, respectively. $\dot{q}_{evp} = \dot{m}_{evp} I$ is the heat loss rate by evaporation, where $\dot{m}_{evp}$ is the mass loss rate by evaporation and calculated with our previous evaporation model \cite{wang2020evaporation}.
\section{Results and discussion}
\label{S:3}
The laser parameters and ambient pressures in the simulation cases are listed in Table\;\ref{tab:Simulation_Paramters}, which are the same as the experiments \cite{zhao2020critical} except the Case 5 (the in-situ experiment under low ambient pressure was not conducted).
The laser spot size is $ 100\;\mathrm{\mu m}$ for all the simulations, the same as that in the experiments \cite{zhao2020critical}. 
The physical properties of Ti-6Al-4V are listed in Table\;\ref{tab:thermal-CFD-values}.
To rule out the influence of the powder particles, all the simulations are conducted on bare plates.
The mesh size is 4 $\mathrm{\mu m}$ to ensure numerical accuracy.
To compare the keyhole pore features, the physical time of all the simulation cases is $2000\;\mathrm{\mu s}$.

\begin{table}[H]
\centering
\caption{Laser scanning parameters and ambient pressures in the simulations}
\label{tab:Simulation_Paramters}
\resizebox{\textwidth}{!}{%
\begin{tabular}{c c c c}
\hline
& Laser power (W) & Laser scanning speed ($\mathrm{mm/s}$) & Ambient pressure (atm) \\ \hline
Case 1 & 365 & 400 & 1.0 \\ 
Case 2 & 382 & 525 & 1.0 \\ 
Case 3 & 382 & 500 & 1.0 \\ 
Case 4 & 382 & 475 & 1.0 \\ 
Case 5 & 382 & 500 & $1.0\times10^{-4}$ \\ 
\hline
\end{tabular}%
}
\end{table}

Based on our simulation results, the pore formation by keyhole collapse has two distinct stages: (i) the instant bubble formation due to the keyhole instability; (ii) the instant bubble pinning on the solidification front.
The keyhole pore formation stages are validated by the X-ray imaging results \cite{zhao2020critical}. 
Although the bubble can be formed due to the keyhole instability, both the simulation and experimental results show that the keyhole instability cannot ensure the instant bubble being captured by the solidification front and forming a pore in the scanning track.

\subsection{Instant bubble due to the keyhole instability}
\label{S:3_1}
\begin{figure}[htb!]
	\centering\includegraphics[width=0.84\linewidth]{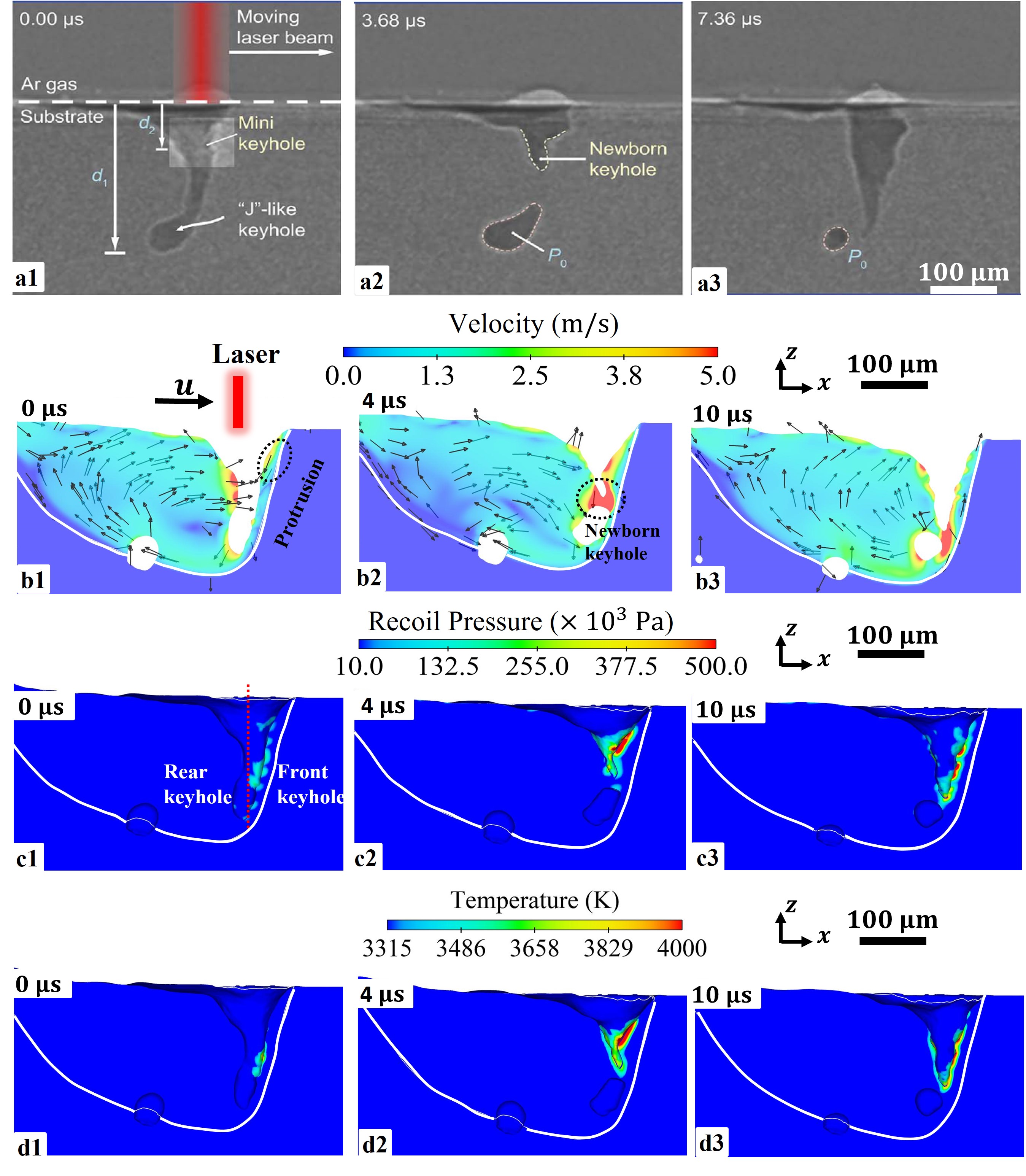}
	\caption{Instant bubble formation due to the keyhole instability. Series a is the X-ray imaging results of the keyhole instability \cite{zhao2020critical}. Series b, c, and d are the velocity, recoil pressure, and keyhole surface temperature in the simulation Case 1. The laser position and scanning direction are indicated in a1 and b1. The arrows in series b represent the velocity directions. The black dashed circle in b1 shows the protrusion on the front keyhole wall. The black dash circle in b2 shows the newborn keyhole. The red dashed line in c1 represents the laser position and separates the keyhole as rear and front part. The white solid lines in b, c, and d series are the contour of the solidus temperature $T_s$. The boiling temperature of Ti-6Al-4V is 3315 K under common ambient pressure. (Experimental figures are from \cite{zhao2020critical}, reprinted with permission from AAAS.)}
	\label{fig:Instability}
\end{figure}

Simulation Case 1 is a validation case on the instant bubble formation process due to the keyhole instability.
According to the X-ray imaging results by \citet{zhao2020critical}, there are four phases of the instant bubble formation by the keyhole fluctuation:
\begin{enumerate}[label=(\arabic*)]
    \item A mini-keyhole protrudes on the top of the front keyhole wall with a letter ``J'' shape keyhole bottom (Fig.\;\ref{fig:Instability} a1).
    \item An instant bubble is formed with the keyhole collapse (Fig.\;\ref{fig:Instability} a2).
    \item The recoil pressure increases with the formation of a new keyhole. A needle-like keyhole bottom (NKB) is formed as the keyhole drills down (Fig.\;\ref{fig:Instability} a3).
    \item The instant bubble moves away from the keyhole.
\end{enumerate}
To compare the keyhole dynamics between the X-ray imaging and simulation results, the starting time point is taken as the mini-keyhole (protrusion) generation time (Fig.\;\ref{fig:Instability} a).
The geometrical feature of the keyhole in the simulation is similar to the experimental results, as presented in Fig.\;\ref{fig:Instability} b and c series in time sequence.
The depths of the newborn keyhole and NKB below the substrate in the experiment are about $99\;\mathrm{\mu m}$ and $226\;\mathrm{\mu m}$ respectively, while they are $115\;\mathrm{\mu m}$ and $150\;\mathrm{\mu m}$ in the simulation as shown in Fig.\;\ref{fig:Instability} b2 and b3.

The velocity field of the molten pool to form the instant bubble is clearly described in Fig.\;\ref{fig:Instability} b series.
In the $1^\mathrm{st}$ phase, the velocity around the middle of the rear keyhole (the rear and front keyhole are defined in Fig.\;\ref{fig:Instability} c1) is higher than other regions of the molten pool as shown in Fig.\;\ref{fig:Instability} b1.
A protrusion, similar to the mini keyhole in Fig.\;\ref{fig:Instability} a1, on the front wall of the keyhole is generated in Fig.\;\ref{fig:Instability} b1, and its velocity increases from the $1^\mathrm{st}$ phase to the $2^\mathrm{nd}$ phase (Fig.\;\ref{fig:Instability} b1 to b2) along the negative $z$ direction so that it finally merges with the rear keyhole wall to form the newborn keyhole.
The high speed at the bottom of the newborn keyhole drives the keyhole penetration further to form a NKB in the $3^\mathrm{rd}$ phase as shown in Fig.\;\ref{fig:Instability} b3.
The distribution of the recoil pressure on the keyhole surface presented in Fig.\;\ref{fig:Instability} c series can further explain the instant bubble formation.
The temperature on the keyhole surface above the boiling temperature is shown in Fig.\;\ref{fig:Instability} d series, where the evaporation on the keyhole surface is a local process.
Correspondingly, there is nearly no recoil pressure on the rear keyhole wall (Fig\;\ref{fig:Instability} c1), which causes the collapse of the keyhole.
If the surface tension, gravity of liquid, hydrodynamic pressure on the rear keyhole is unbalanced, the keyhole would collapse like Fig.\;\ref{fig:Instability} c2.
When the newborn keyhole wall forms, the evaporation in Fig.\;\ref{fig:Instability} d2 becomes much more violent and the recoil pressure drives the keyhole tip deeper. 
Thus, the unevenly distributed recoil pressure is the driving force for the keyhole collapse and penetration.

\citet{zhao2020critical} proposed that in the $4^\mathrm{th}$ phase a shock wave by the phase explosion \cite{zhao2019bulk,miotello1999laser} causes the keyhole tip to penetrate the bubble $\mathbf{P_0}$ and splitting it into two bubbles.
It should be clarified that the sharp shock wave cannot be simulated by the current model, because the phase explosion is a fierce non-linear process that occurs in less than $1\;\mathrm{\mu s}$ with ``water hammer'' pressure on $\mathbf{P_0}$ at $\sim 140\;\mathrm{MPa}$.
More importantly, the shock wave is not the decisive factor for the pinning of the bubble at the solidification front to form the keyhole pore.
As the bubble is nearly stationary after its formation, the shock wave triggers and splits the bubble into two, which subsequently escapes to the surface of the molten pool. 
Hence, no keyhole pore forms when a shock wave is encountered. 
Therefore, more simulation cases are conducted to investigate how the instant bubble pins on the solidification front and eventually forms the keyhole pores as observed in experiments \cite{zhao2020critical}.

Before further discussion of the keyhole pore formation, the issue below should be clarified.
In the simulation, there is a keyhole pore with the diameter of $45\;\mathrm{\mu m}$ by two merged bubbles as shown in Fig.\;\ref{fig:Instability} b3, while there is no keyhole pore in the experiment \cite{zhao2020critical}.
There could be several reasons for this difference.
The experimental results by \citet{zhao2020critical} show that the laser parameters in Case 1 are close to threshold for no keyhole pore, where the possibility of the keyhole pore formation is around zero.
Meanwhile, the diameter of the laser in the experiment is an approximate value, and the influence of laser defocusing and plasma are not incorporated in the simulations.
Thus, this kind of random error is acceptable.
\subsection{Keyhole pore formation}
\label{S:3_2}
\begin{figure}[htb!]
	\centering\includegraphics[width=\linewidth]{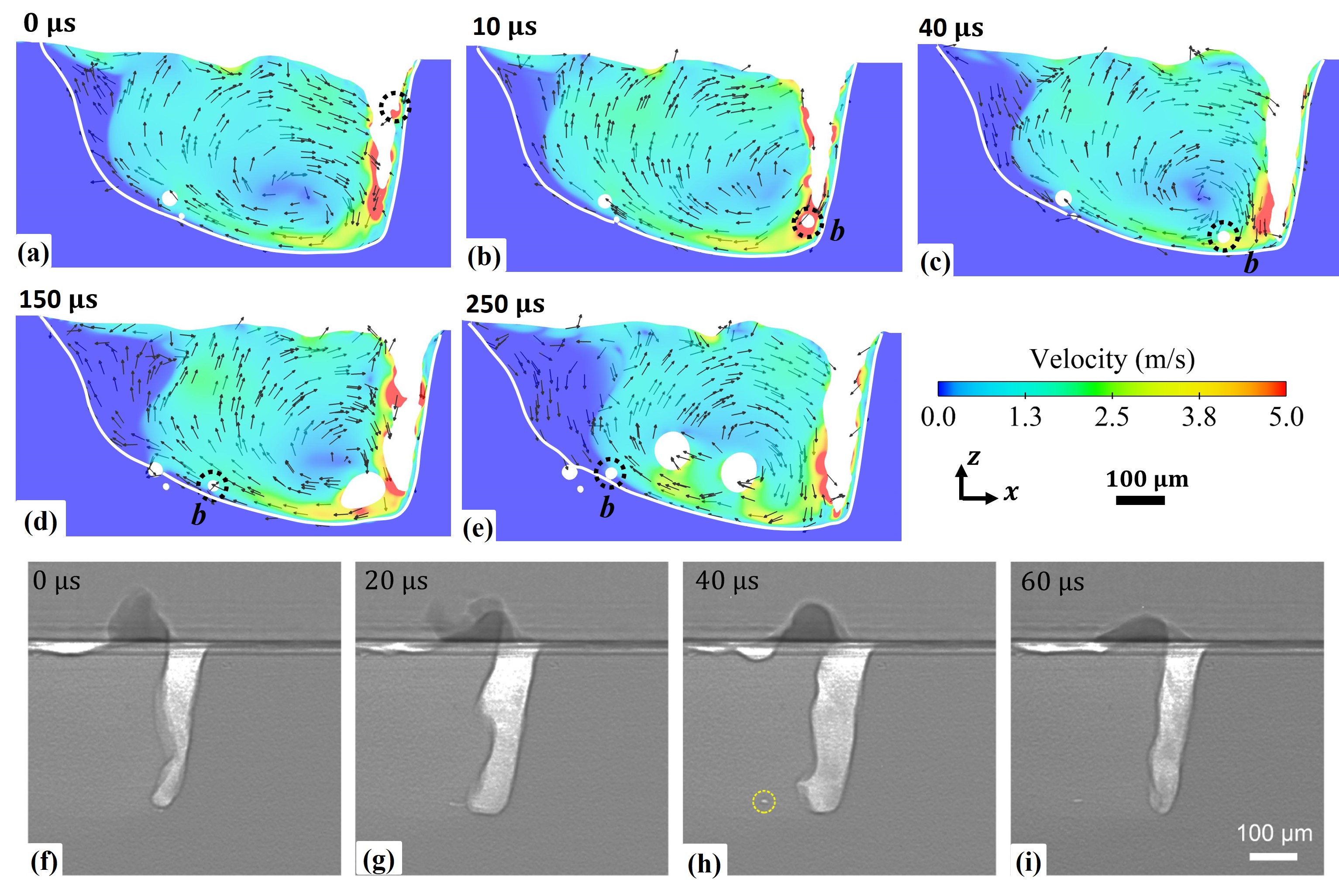}
	\caption{Keyhole pore formation in Case 2. (a-e) The simulation results of the velocity magnitude and direction (arrows) in the molten pool in the longitudinal section (2D views). (f-i) The keyhole pore formation captured by X-ray imaging \cite{zhao2020critical}. The black dashed circle in (a) is the mini keyhole. The bubble $\mathbf{b}$ in (b-e) is the instant bubble by the keyhole collapse. The white solid lines in (a-e) are the contour of the solidus temperature $T_s$. The instant bubble in the experiment is shown in the yellow circle of (h). (Experimental figures are from \cite{zhao2020critical}, reprinted with permission from AAAS.)}
	\label{fig:poreFormation}
\end{figure}

According to the experimental results \cite{zhao2020critical}, instant bubbles pin on the solidification front and form the keyhole pores under the manufacturing parameters in Case 2-4.
The simulation result of Case 2 is taken as an example to explain the keyhole pore formation process, and the velocity distribution in the molten pool during the keyhole pore formation is shown in Fig.\;\ref{fig:poreFormation} (a-e).
In Fig.\;\ref{fig:poreFormation} (a) and (b), the instant bubble $\mathbf{b}$ is formed as explained in Sect.\;\ref{S:3_1}.
A protrusion is formed on the front keyhole wall as shown in the black dashed circle of Fig.\;\ref{fig:poreFormation} (a), and an instant bubble $\mathbf{b}$ is generated by the keyhole collapse in Fig.\;\ref{fig:poreFormation} (b).
It takes about $240\;\mathrm{\mu s}$ for the instant bubble $\mathbf{b}$ to be pinned on the solidification front (Fig.\;\ref{fig:poreFormation} (e)).

The bubble motion and pore formation at the bottom of the molten pool in both the experiment and simulation are similar during the instant bubble pinning stage.
The bubble $\mathbf{b}$ mainly moves along the horizontal direction without floating up as shown in Fig.\;\ref{fig:poreFormation} (c) and (d), similar to the experimental result in Fig.\;\ref{fig:poreFormation} (g) and (h).
About $20\sim30\;\mathrm{\mu s}$ after the keyhole collapse, the horizontal distance between the bubble $\mathbf{b}$ and the keyhole bottom is about $112\;\mathrm{\mu m}$ and $129\;\mathrm{\mu m}$ in the simulation and experiment, respectively, as shown in Fig.\;\ref{fig:poreFormation} (c) and (h).
Furthermore, by observing the velocity distribution around the instant bubble $\mathbf{b}$, we find that the liquid velocity below the bubble is much higher than that above in Fig.\;\ref{fig:poreFormation} (c).
Based on the Bernoulli's principle, the high velocity below the pore leads to low local pressure, which impedes the upward motion of the bubble.
Eventually the bubble is caught up by the solidification front.
Thus, the velocity distribution at the bottom of the molten is one of the factors for the formation of the keyhole pore.
This phenomenon also explains larger bubbles bear larger buoyancy and thus float up from the molten pool bottom with less hindrance in Fig.\;\ref{fig:poreFormation} (e).

\begin{figure}[H]
	\centering\includegraphics[width=0.8\linewidth]{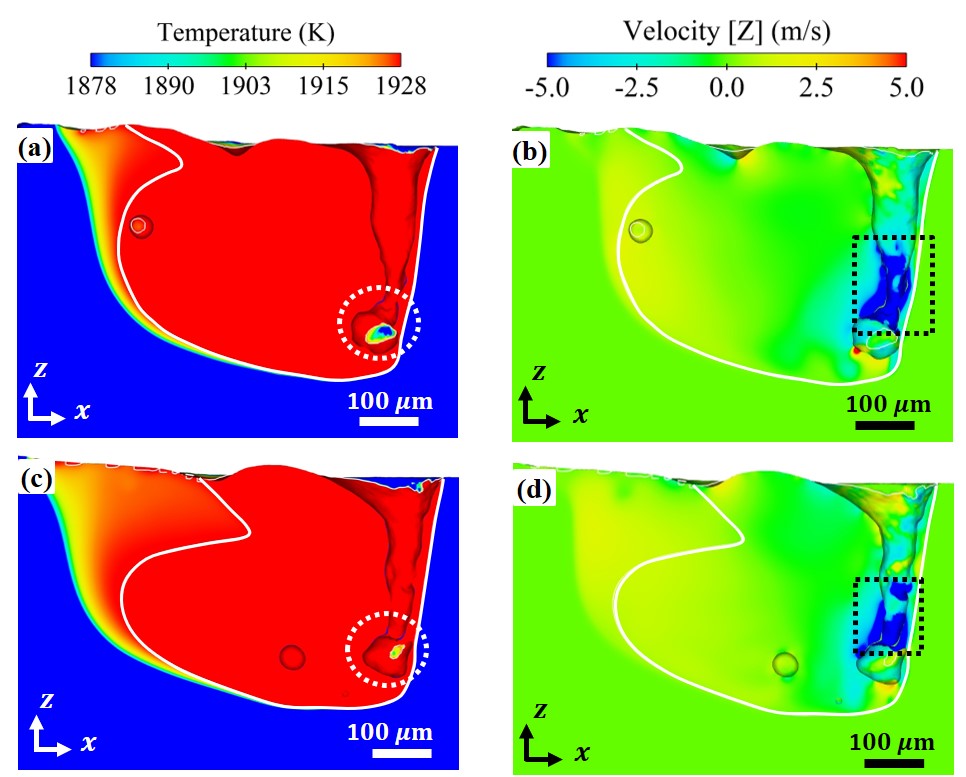}
	\caption{Influence of mushy zone on the keyhole fluctuation. (a and b) are the temperature and $z$-direction velocity distribution in the molten pool at $t=1134\;\mathrm{\mu s}$. (c and d) are the temperature and $z$-direction velocity distribution in the molten pool at $t=1302\;\mathrm{\mu s}$. The white solid lines in (a-d) are the contour of the liquidus temperature $T_l$.}
	\label{fig:Darcy}
\end{figure}

Additionally, the drag force in the mushy zone is acting on the fluctuating keyhole, especially at the keyhole bottom.
The interaction between the mushy zone and keyhole surface does not happen coincidentally.
The molten pool temperature fields at two different time points in the $1^\mathrm{st}$ phase of the keyhole collapse in simulation Case 2 are given in Fig.\;\ref{fig:Darcy} (a) and (c).
The regions in the white dashed circle of Fig.\;\ref{fig:Darcy} (a) and (c) indicate that the keyhole surface contacts the mushy zone at different times.
According to Eq.\;\ref{eq:darcy}, the drag force in the mushy zone increases hyperbolically as the solid fraction increases, and the drag coefficient $D$ is larger than $1.2\times10^6\;\mathrm{s^{-1}}$ when the solid fraction is larger than 0.25.
Thus, the front part of the keyhole bottom can hardly expand, while the rear part of the keyhole bottom expands to form a ``J'' shape keyhole, as the metal evaporates.
Moreover, the higher amplitude of $z$ direction velocity around the middle of the molten pool accelerates the keyhole collapse as shown in the dashed rectangles of Fig.\;\ref{fig:Darcy} (b) and (c).
It should be mentioned that as the grain morphology in the mushy zone is mostly columnar and rarely equiaxed \cite{debroy2018additive,raghavan2016numerical,yang2021phase}, the drag force should be anisotropic.
The drag force by the Blake-Kozeny model used here is isotropic.
Thus, a more accurate Darcy drag force model with the consideration of anisotropy of the mushy zone morphology is worthy of investigation in future study.

\begin{figure}[H]
	\centering\includegraphics[width=0.7\linewidth]{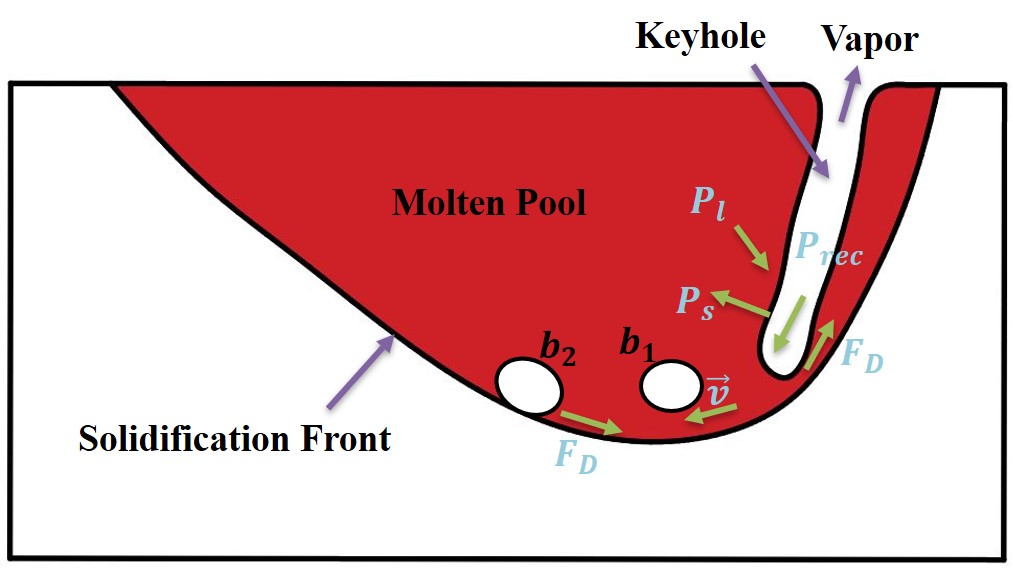}
	\caption{Schematic of the keyhole pore formation process. The recoil pressure ($\boldsymbol{P}_{rec}$) by metal evaporation, hydrodynamic pressure ($\boldsymbol{P}_l$), pressure by surface tension ($\boldsymbol{P}_s$), and drag force in the mushy zone ($\boldsymbol{F}_D$) lead to the instability of the keyhole and generate bubbles $\mathbf{b}_1$ and $\mathbf{b}_2$. The bubbles are not floating up directly due to Bernoulli's principle and are caught by the solidification front.}
	\label{fig:poreScheme}
\end{figure}

Based on the analysis of the instant bubble formation and pinning on solidification front stages, the schematic of the keyhole pore formation is presented in Fig.\;\ref{fig:poreScheme}. 
With the unbalanced forces, the keyhole fluctuates up and down and even collapses to form instant bubbles ($\mathbf{b}_1$ and $\mathbf{b}_2$).
In the bubble pinning on solidification front stage, the molten pool bottom has higher velocity than that above the bubble $\mathbf{b}_1$ and lower pressure, which leads to a vertical drag force on the bubble $\mathbf{b}_1$.
If the drag force is sufficiently strong, the bubble $\mathbf{b}_1$ would move nearly horizontally and pins on the solidification front like $\mathbf{b}_2$.

\subsection{Keyhole pore feature}
\label{S:3_3}
\begin{figure}[htb!]
	\centering\includegraphics[width=\linewidth]{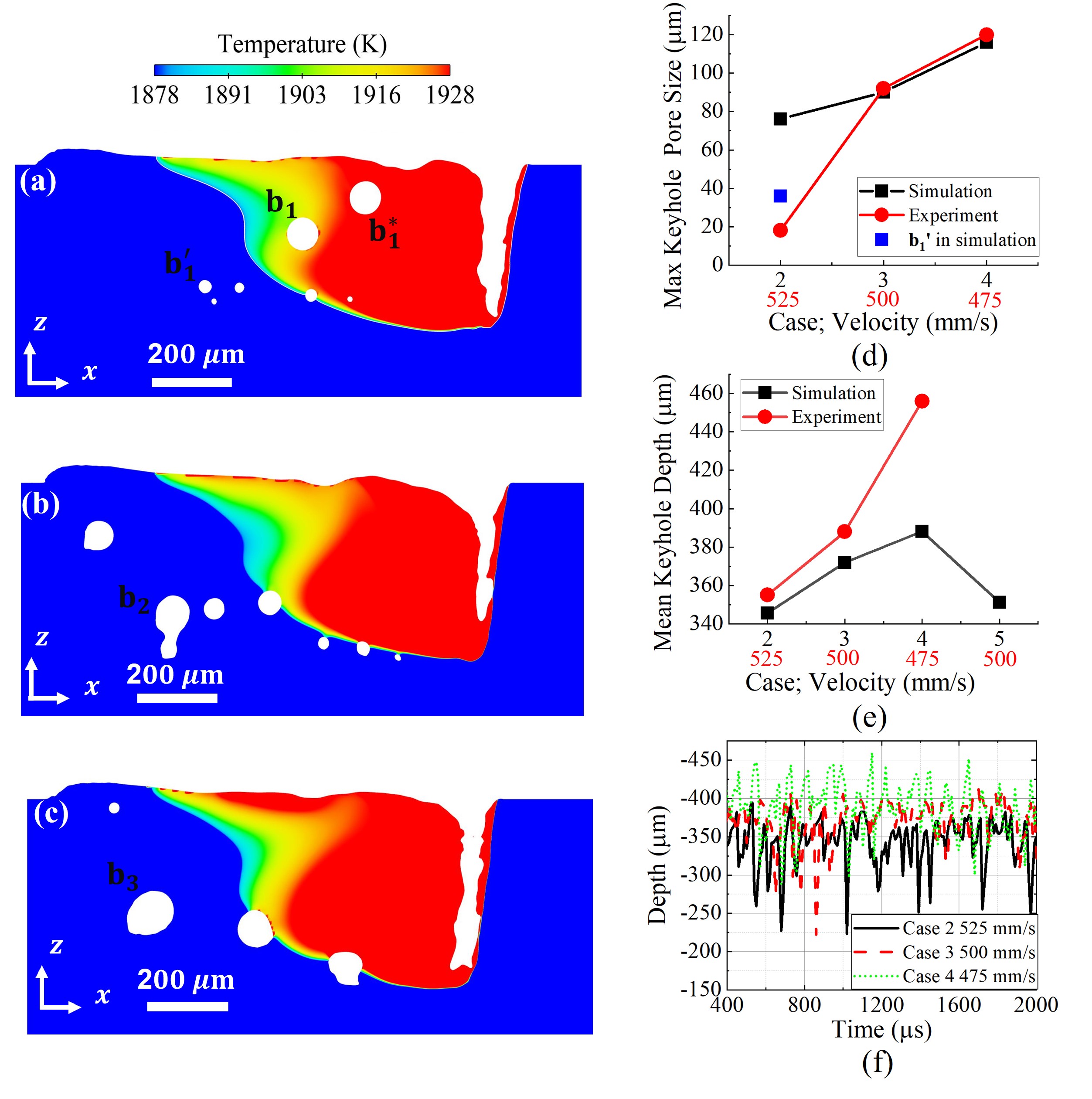}
	\caption{Keyhole and keyhole pore features. 
	Keyhole pores in (a) Case2 ($525\;\mathrm{mm/s}$), (b) Case 3 ($500\;\mathrm{mm/s}$), and (c) Case 4 ($475\;\mathrm{mm/s}$) in the center plane of the scanning track at $t = 2000\;\mathrm{\mu s}$. 
	(d) Maximum keyhole pore size and (e) mean keyhole depth between simulation and experiment.
	(f) Keyhole depth fluctuation with time.
	$\mathbf{b}_1$, $\mathbf{b}_2$, and $\mathbf{b}_3$ are the maximum pore in Case2- 4, respectively. $\mathbf{b}_1'$ in (a) is the second largest pore in simulation Case 2, indicated in (e). The case number and laser scanning speed are shown in (d-e) together for better comparison.}
	\label{fig:poreFeature}
\end{figure}

The keyhole and keyhole pore features in simulation Case 2-4 are presented in Fig.\;\ref{fig:poreFeature} and listed in Table\;\ref{tab:key_depth_pore}.
The maximum keyhole pore sizes in both simulation and experiment decrease as the laser scanning speed decreases (with the same laser power).
At low scanning speeds, the maximum keyhole pore sizes in simulation and experiment is close to each other with differences lower than $4\;\mathrm{\mu m}$ as shown in Fig.\;\ref{fig:poreFeature} (d).
When the laser scanning speed reaches the threshold for no keyhole pore (Case 2), the contingency of the maximum keyhole pore size increases.
In simulation Case 2, the bubble $\mathbf{b}_1$ (size of $76\;\mathrm{\mu m}$) is pinned on the solidification front, while the similar bubble $\mathbf{b}_1^*$ (size of $80\;\mathrm{\mu m}$) floats up to the molten pool surface.
Nonetheless, the size of the second maximum keyhole pore $\mathbf{b}_1'$ is $36\;\mathrm{\mu m}$ and close to the experimental result of $18\;\mathrm{\mu m}$.
It should be mentioned that the keyhole pore size is sensitive to the experimental parameters near the threshold for no keyhole pore situations, such as the laser focusing and defocusing, material purity, and laser deflection of the vapor plume, etc.
While the current simulation model cannot incorporate all the influence factors, the difference of maximum keyhole pore size in Case 2 is acceptable.
Since the maximum keyhole pore size has certain randomness in both simulation and experiment, 
other parameters, such as mean keyhole pore size, the median value of pore size, and pore distribution, would be more representative, but were not provided by the experiment \cite{zhao2020critical}.
The mean keyhole pore size in the simulations increases steadily as the laser scanning speed decreases (Table\;\ref{tab:key_depth_pore}).

The keyhole pores are concentrated at the bottom of the laser scanning track with smooth surfaces compared to the lack-of-fusion pores.
Further observation of the keyhole pore distribution in Fig.\;\ref{fig:poreFeature} (a-c) shows that not only does the keyhole pore size increase sharply as the laser scanning speed decreases, but also the position and shape of the pores change.
In Case 2, the keyhole pores are spherical and  horizontally distributed in the laser scanning track.
As the laser scanning speed decreases, the shape of the keyhole pores becomes increasingly irregular with a flat or sharp bottom, while the upper surface remains smooth and spherical.
Additionally, the different sizes of the instant bubbles in the low scanning speed cases also lead to different buoyant forces, vertical motions to a certain degree and thus non-horizontal distribution of the pores.

\begin{table}[htb!]
\centering
\caption{Keyhole pore size and depth statistical features}
\label{tab:key_depth_pore}
\begin{threeparttable}
\begin{tabular}{m{0.3\textwidth} m{0.08\textwidth} m{0.1\textwidth} m{0.1\textwidth} m{0.1\textwidth} m{0.1\textwidth} }
\hline
& & Case 2 & Case 3 & Case 4  & Case 5 \\ \hline
\multirow{2}{0.32\textwidth}{Mean keyhole depth ($\mathrm{\mu m}$)}   
& Exp.\tnote{a} & 355    & 388    & 456     & -      \\ 
& Sim.\tnote{b} & 345    & 372    & 388     & 351 \\
\hline
Mean keyhole pore size\tnote{c} ($\mathrm{\mu m}$)    & & 37    & 49     & 82 & -\\
\hline
Standard deviation of keyhole depth ($\mathrm{\mu m}$) & & 30    & 29     & 35    & 37\\
\hline
Maximum fluctuation of keyhole depth\tnote{d} ($\mathrm{\mu m}$)    & & 152      & 170      & 156     & 164\\  
\hline
\end{tabular}%

\begin{tablenotes}
    \item[a] Exp. represents the experiment result.
    \item[b] Sim. represents the simulation result.
    \item[c] The pores in the solidified region are considered.
    \item[d] It is the distance from the peak to the valley of keyhole depth curves. 
\end{tablenotes}
\end{threeparttable}
\end{table}

As the instant bubble formation is related to the keyhole collapse, the statistical features of the keyhole depth are given in Fig.\;\ref{fig:poreFeature} (e-f) and Table\;\ref{tab:key_depth_pore}.
The mean keyhole depths in the simulation cases match those in the experiments (Fig.\;\ref{fig:poreFeature} (e)), although the relative difference of the keyhole depth between the simulations and experiments increases from 2.5\% to 15\% as the laser scanning speed decreases.
This diverging discrepancy can be explained by several reasons.
Firstly, the aforementioned bulk explosion is fiercer as the energy density increases.
Next, the plume in a deeper keyhole traps more photons, leading to greater energy absorptivity in the material. 
This effect is not considered in the current model.
Additionally, the laser diameter varies along the keyhole depth due to defocusing, which influences the laser reflections and energy absorption. The laser in the current model is parallel with a constant diameter.

As the laser scanning speed decreases, the keyhole depth curves in Fig.\;\ref{fig:poreFeature} (f) show an obvious increase, similar to the increasing trend of mean keyhole pore size.
Moreover, the standard deviation of the keyhole depth in Case 4 are obviously larger than those Case 2 and 3 in Table\;\ref{tab:key_depth_pore}, which indicates that the keyhole fluctuation in Case 4 is much more violent.
Compared to the mean keyhole depth, the maximum fluctuations of keyhole depth in three cases are larger than 40\% of the mean keyhole depth. 
The analysis of Sect.\;\ref{S:3_1} indicates that the higher keyhole depth variation between the $1^\mathrm{st}$ phase and the $2^\mathrm{nd}$ phase makes the keyhole more unstable and generates instant bubbles more easily.
Thus, there is a higher probability of the keyhole pore formation in Case 4.
Additionally, the X-ray imaging results \cite{hojjatzadeh2020direct,zhao2020critical} also show that the shallow keyhole with a large keyhole fluctuation distance can lead to pores, especially in the case of laser scanning the powder bed (the spattering particles influence  laser absorption and metal vapor flow).
It suggests that the keyhole depth and its fluctuation are two important parameters for keyhole pore formation and should be considered for future observations and analysis.

\begin{figure}[htb!]
	\centering\includegraphics[width=\linewidth]{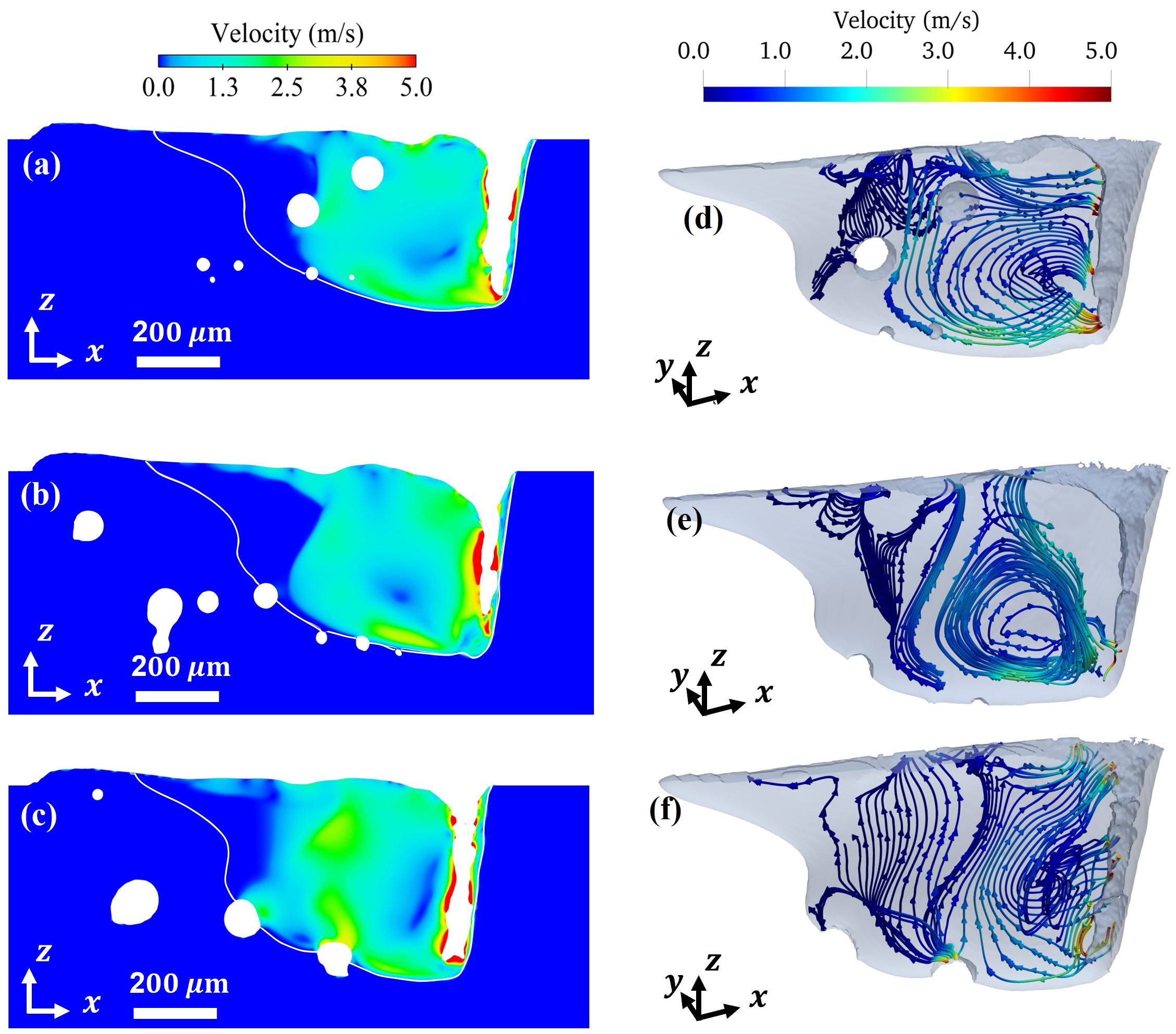}
	\caption{(a-c) Velocity magnitude and (d-f) streamline in the molten pool at $t = 2000\;\mathrm{\mu s}$ in Case2 (a and d), Case 3 (b and e), and Case 4 (c and f). The white solid curves in (a-c) and gray contours in (d-f) are the contour of the solidus temperature $T_s$.}
	\label{fig:streamline}
\end{figure}

The velocity magnitudes and streamlines of the Case 2-4 share similar distributions as plotted in Fig.\;\ref{fig:streamline}.
The velocity magnitude is larger around the keyhole and molten pool bottom than other regions.
The streamline distributions in Fig.\;\ref{fig:streamline} (d-f) show that there are two vortices in the molten pool: a larger stronger clockwise vortex at the front part of the molten pool and a smaller weaker anticlockwise vortex at the rear part of the molten pool. 
The distribution and location of the vortex pair match well with those observed in the X-ray imaging experiments by \citet{hojjatzadeh2019pore}.
As the keyhole pores always form at the bottom of the melt track, the velocity distribution at the bottom of the melt pool is more relevant and influential.
At the molten pool bottom, the liquid flows horizontally to the rear part of the molten pool and thus drives the instant bubble backward to the solidification front. 


\begin{figure}[htb!]
	\centering\includegraphics[width=\linewidth]{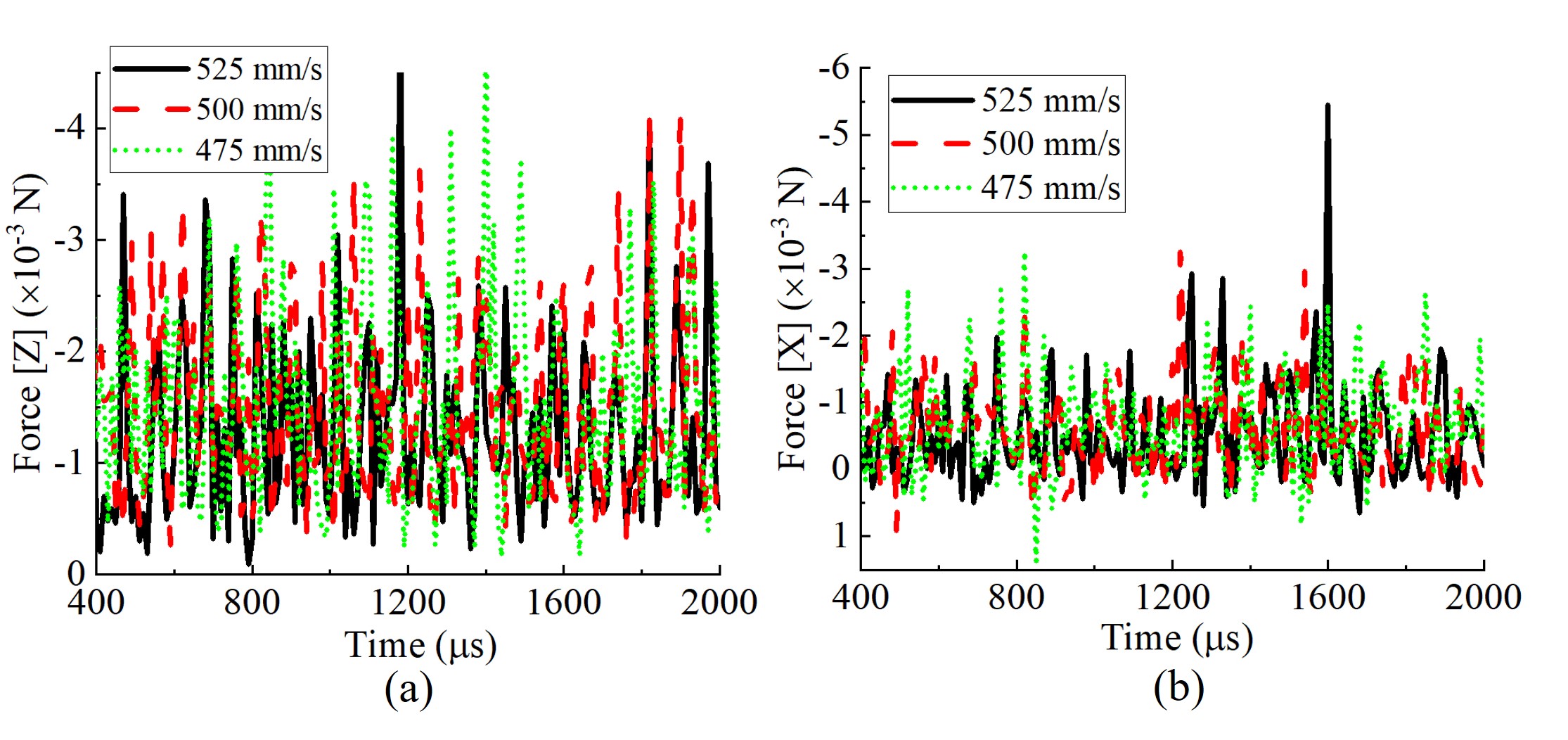}
	\caption{Recoil forces on the keyhole surface in (a) $z$ and (b) $x$ directions in Case2 ($525\;\mathrm{mm/s}$), Case 3 ($500\;\mathrm{mm/s}$), and Case 4 ($475\;\mathrm{mm/s}$). The keyhole is nearly symmetry along the $y$ direction. Thus, the recoil force along the $y$ direction is around 0 and not plotted.}
	\label{fig:recoilForce}
\end{figure}

In Case 2-4, the total recoil force, i.e., the integral of recoil pressure over the keyhole surface, acts mainly along the $z$ direction as shown in Fig.\;\ref{fig:recoilForce}, and the $z$ components of the recoil forces in these cases have higher amplitudes and are fluctuating around $-1\sim-4\times10^{-3}\;\mathrm{N}$, compared to the $x$ components fluctuating around $0\sim-2\times10^{-3}\;\mathrm{N}$.
However, it is difficult to tell the difference of the recoil forces in the three cases.
This means that the keyhole instability and keyhole pore formation are more related to the locally distributed recoil force rather than the total force on the whole keyhole surface.

\begin{figure}[htb!]
	\centering\includegraphics[width=\linewidth]{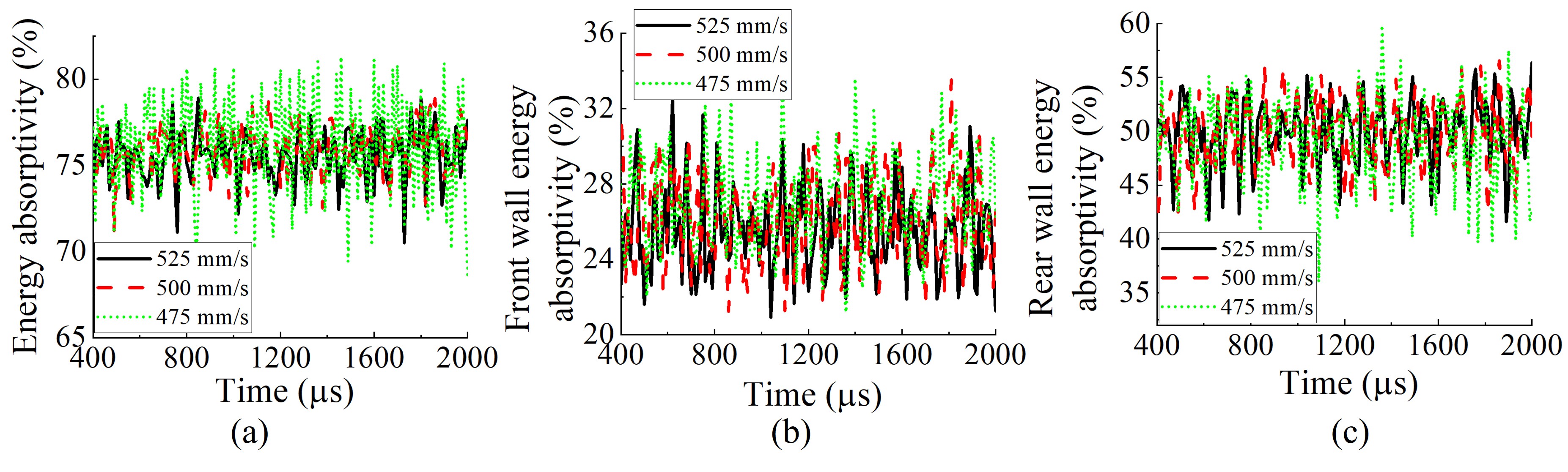}
	\caption{Energy absorptivity on (a) the whole keyhole surface, (b) front keyhole  wall, and (c) rear keyhole wall in Case2 ($525\;\mathrm{mm/s}$), Case 3 ($500\;\mathrm{mm/s}$), and Case 4 ($475\;\mathrm{mm/s}$).}
	\label{fig:absorptivity}
\end{figure}

Further analysis of the energy absorptivity in the three cases hints the relationship between the keyhole pore formation and keyhole dynamics.
The total energy absorptivity and that on rear keyhole wall are different from that on the front keyhole wall (Fig.\;\ref{fig:absorptivity}).
In simulation Case 4, the total energy absorptivity and that on rear keyhole wall have fiercer fluctuations compared to that on the front keyhole wall.
As the laser reflection is determined by the morphology of the keyhole, it indicates that the shape variation of the rear keyhole wall is larger in Case 4.
The mean value and standard deviation of the absorptivity are listed in Table\;\ref{tab:absorptivity}.
While the mean absorptivity on each part of the keyhole surface is similar in the three cases, the standard deviation on the rear keyhole wall increases as the scanning speed decreases.
This observation implies that the keyhole morphology is sensitive to the laser scanning speed, especially the rear keyhole wall, which is closely related to the keyhole fluctuation and instant bubble formation.
In other words, the energy absorptivity fluctuation could be an effective criterion to evaluate the possibility of the keyhole pore formation.

The above analysis of the keyhole pore features in Case 2-4 shows that the keyhole pore size is sensitive to the manufacturing parameters. 
The increase of laser scanning speed is $50 \;\mathrm{mm/s}$, while the mean keyhole pore size decreases from $82\;\mathrm{\mu m}$ to $37\;\mathrm{\mu m}$.
While the molten pool flow of the cases are similar, the deviation of keyhole depth and energy absoptivity fluctuation reflect the keyhole instability.
\begin{table}[H]
\centering
\caption{Energy absorptivity on the front and rear keyhole wall}
\label{tab:absorptivity}
\begin{threeparttable}

    \begin{tabular}{cccccc}
    \hline
    \multicolumn{2}{c}{Energy absorptivity}  & Case 2 & Case 3 & Case 4 & Case 5\\
    \hline
    \multicolumn{1}{c}{\multirow{2}{*}{Front keyhole wall}} & Mean (\%) & 25.54  & 26.07  & 27.07 & 25.58\\
    \multicolumn{1}{c}{}    & SD\tnote{a} (\%)      & 2.29   & 2.51   & 2.63  & 2.23\\
    \hline
    \multirow{2}{*}{Rear keyhole wall}  & Mean (\%) & 50.04  & 49.88  & 49.04  & 50.40\\
    & SD\tnote{a} (\%)      & 3.02   & 3.26   & 4.21  & 2.75\\
    \hline
    \end{tabular}%

    \begin{tablenotes}
        \item[a] SD means standard deviation.
    \end{tablenotes}
\end{threeparttable}

\end{table}

\subsection{Reduce keyhole pores: low ambient pressure}
The mechanisms of keyhole pore formation as discussed above shows that the balance of the forces on the keyhole surface is important to ameliorate the pores formed by keyhole fluctuation.
Although it has been proven that the porosity of the as-built part under near-vacuum and low ambient pressure can be reduced during both AM and laser welding \cite{zhang2013microstructure,zhou2018study,jiang2020mitigation}, the principles on how the ambient environment influences the keyhole and molten pool dynamics and reduce the pores during AM are rarely studied yet.
To explore the principles of using low ambient pressure to reduce keyhole pores, a simulation (Case 5) is conducted under $10^{-4}\;\mathrm{atm}$, in which the other meanufacturing parameters are kept the same as Case 3.

\begin{figure}[htb!]
	\centering\includegraphics[width=\linewidth]{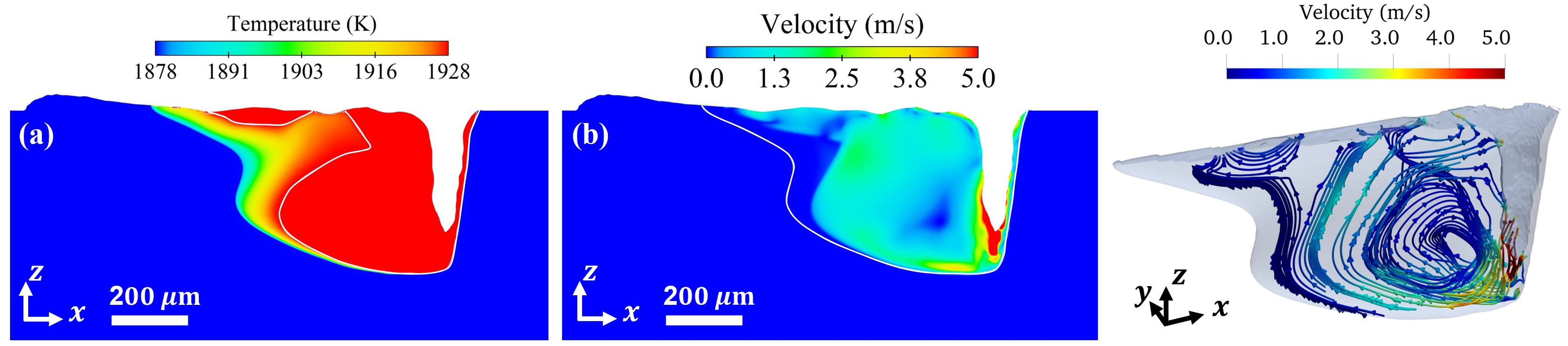}
	\caption{Simulation results of the molten pool flow under $10^{-4}\;\mathrm{atm}$ ambient pressure (Case 5) at $t=2000\;\mathrm{\mu s}$. (a) Temperature distribution, (b) velocity magnitude distribution, and (c) streamline in the molten pool. The white curve in (a) is the contour of the liquidus temperature $T_l$, and the white curve in (b) and grey contour in (c) are the contour of the solidus temperature $T_s$.}
	\label{fig:10Pa}
\end{figure}

As presented in Fig.\;\ref{fig:10Pa}, there are no keyhole pores in the simulation domain in Case 5.
The decreasing trend of pore formation under low ambient pressure in the simulation is similar to the experiments \cite{zhou2018study,katayama2001effect}, where there are nearly no pores under the ambient pressure below $100\;\mathrm{Pa}$.
The temperature, velocity distribution, and streamline in Case 5 are similar to those in Case 3.
Despite the decrease in ambient pressure, the molten pool shapes are similar, and the depth increases slightly from $416\;\mathrm{\mu m}$ to $430\;\mathrm{\mu m}$.
A vortex pair is observed, with a larger and stronger clockwise rotating vortex at the front part of the molten pool. 
At the bottom region, the velocity decreases longitudinally from the front to the rear end of the molten pool.

\begin{figure}[htb!]
	\centering\includegraphics[width=\linewidth]{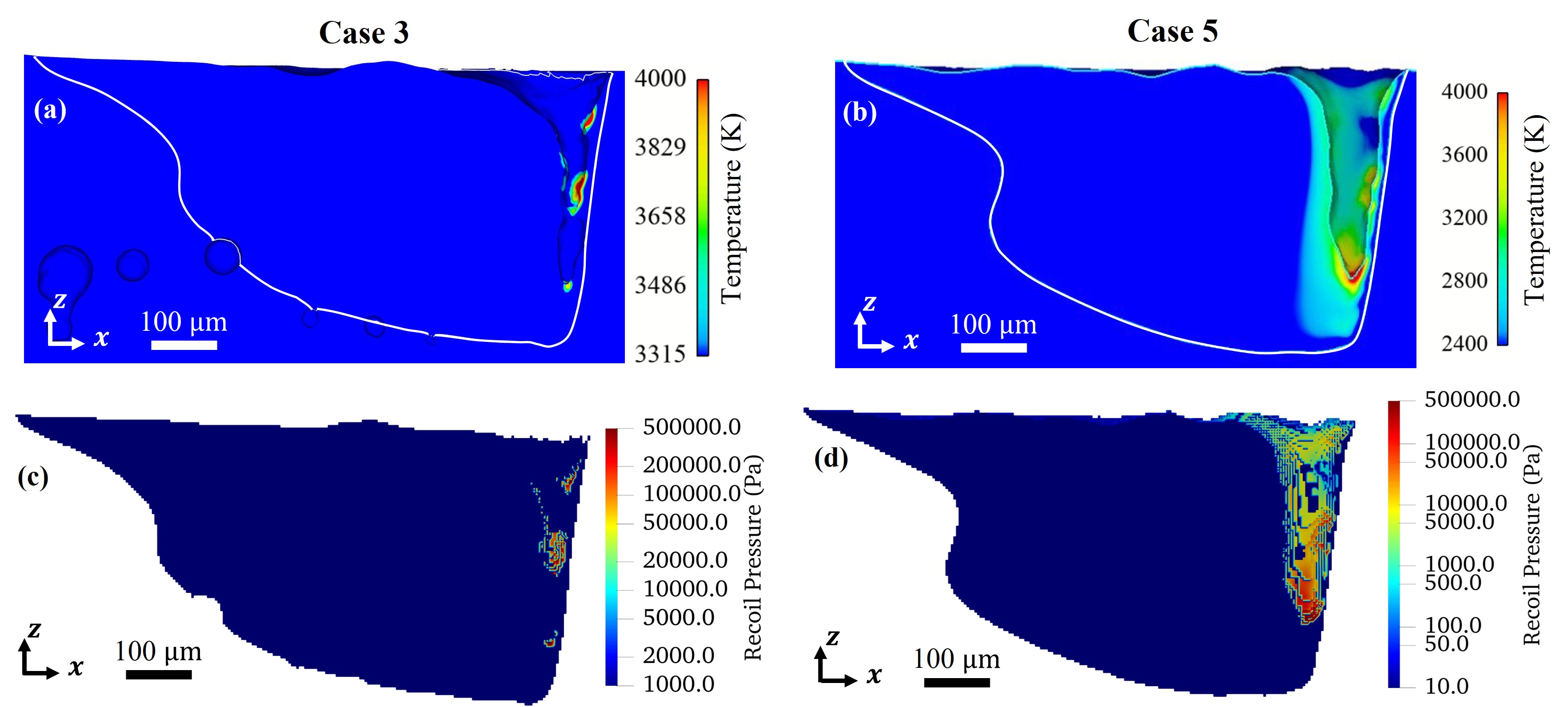}
	\caption{(a and b) Temperature field above the boiling temperatures and (c and d) recoil pressure  on the keyhole surface under the common ambient (Case 3) and low ambient pressure (Case 5). (a and c) and (b and d) are the simulation results of Case 3 and 5, respectively. The white curves in (a) and (b) are the contour of the solidus temperature $T_s$.}
	\label{fig:recDiff}
\end{figure}

\begin{figure}[htb!]
	\centering\includegraphics[width=0.6\linewidth]{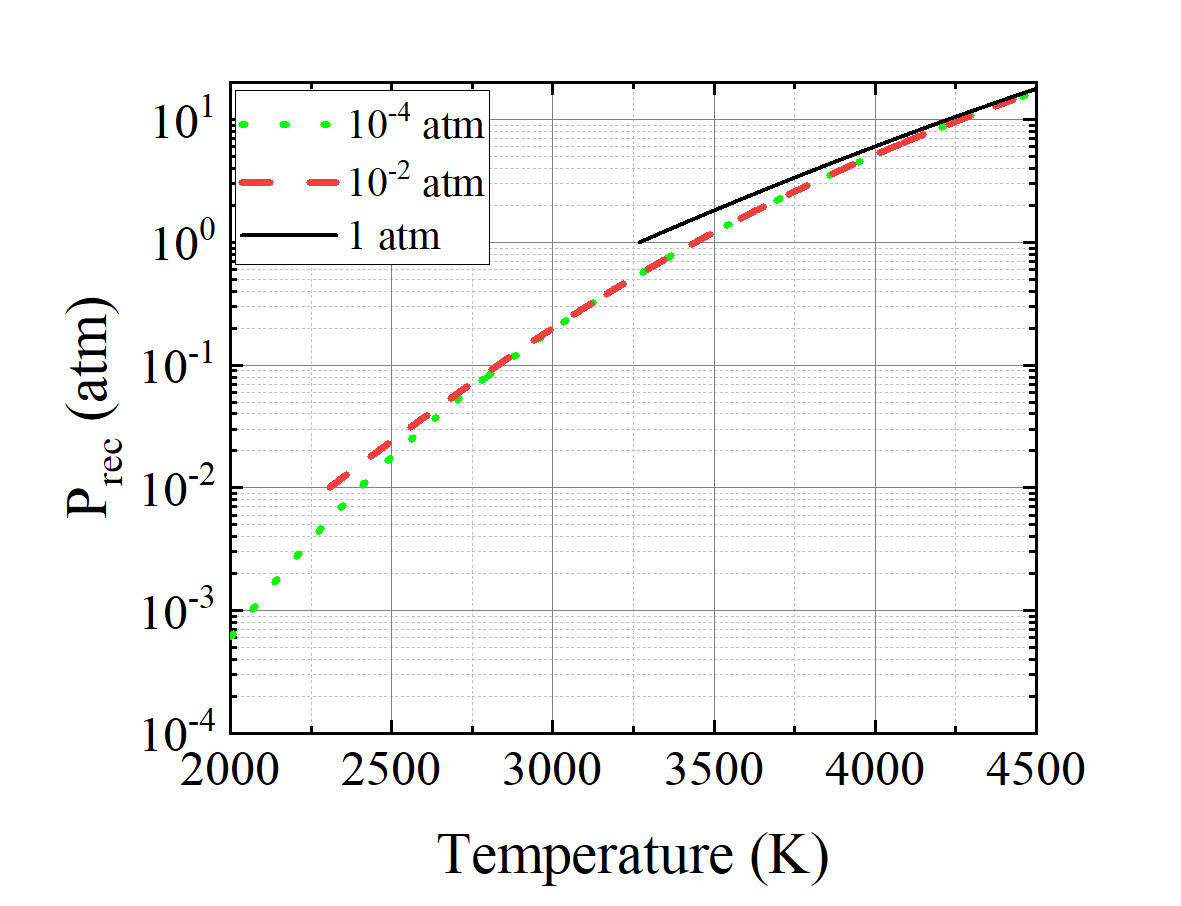}
	\caption{Relationship between recoil pressure and surface temperature of Ti-6Al-4V  under different ambient pressures. The boiling temperature varies with the ambient pressure, so the starting points of the three curves are different.}
	\label{fig:rec-Temperature}
\end{figure}

The main differences between Case 3 and 5 are the ambient pressure and consequently the recoil pressure.
The keyhole fluctuation is dynamic, and the recoil pressure changes with the ambient pressure and keyhole surface temperature. 
Thus, further analysis of the recoil pressure on the keyhole surface is necessary as given in Fig.\;\ref{fig:recDiff}.
Calculated by our evaporation model \cite{wang2020evaporation}, the recoil pressure-surface temperature curves under different ambient pressures are different as shown in Fig.\;\ref{fig:rec-Temperature}.
Although the amplitude of the recoil pressure at high temperature is similar, the boiling temperature decreases as the ambient pressure drops.
Therefore, the temperature and recoil pressure ranges in Fig.\;\ref{fig:recDiff} for the two cases are different.

In Case 3, the region near the keyhole with the temperature above the boiling point is small and randomly distributed on the front keyhole wall (Fig.\;\ref{fig:recDiff} (a)).
Correspondingly, the recoil pressure is  concentrated on the front keyhole wall (Fig.\;\ref{fig:recDiff} (c)).
Thus, the rear keyhole wall has weak support from the recoil pressure and collapses easily to form instant bubbles.
In contrast, the recoil pressure distribution under the low ambient pressure is different.
Although the highest temperature on the keyhole surface in Fig.\;\ref{fig:recDiff} (b) is close to that in Fig.\;\ref{fig:recDiff} (a) at around 4000 K, the temperature of nearly entire keyhole surface is above the boiling temperature and the temperature increases continuously from the upper to the bottom region of the keyhole.
This temperature distribution indicates stronger evaporation from keyhole surface under the low ambient pressure, which follows the trend in previous experiments \cite{zhou2018study,zhou2019research}.
Moreover, the recoil pressure on the keyhole surface is distributed all over the keyhole surface with the value above 0.1 atm and up to 5 atm at the bottom of the keyhole.
Compared to the recoil pressure distribution in Case 3 (Fig.\;\ref{fig:recDiff} (c)), the recoil pressure on the rear keyhole wall under the low ambient pressure is higher which stabilizes the rear keyhole wall during fluctuation and thus reduces keyhole pores.

\begin{figure}[H]
	\centering\includegraphics[width=\linewidth]{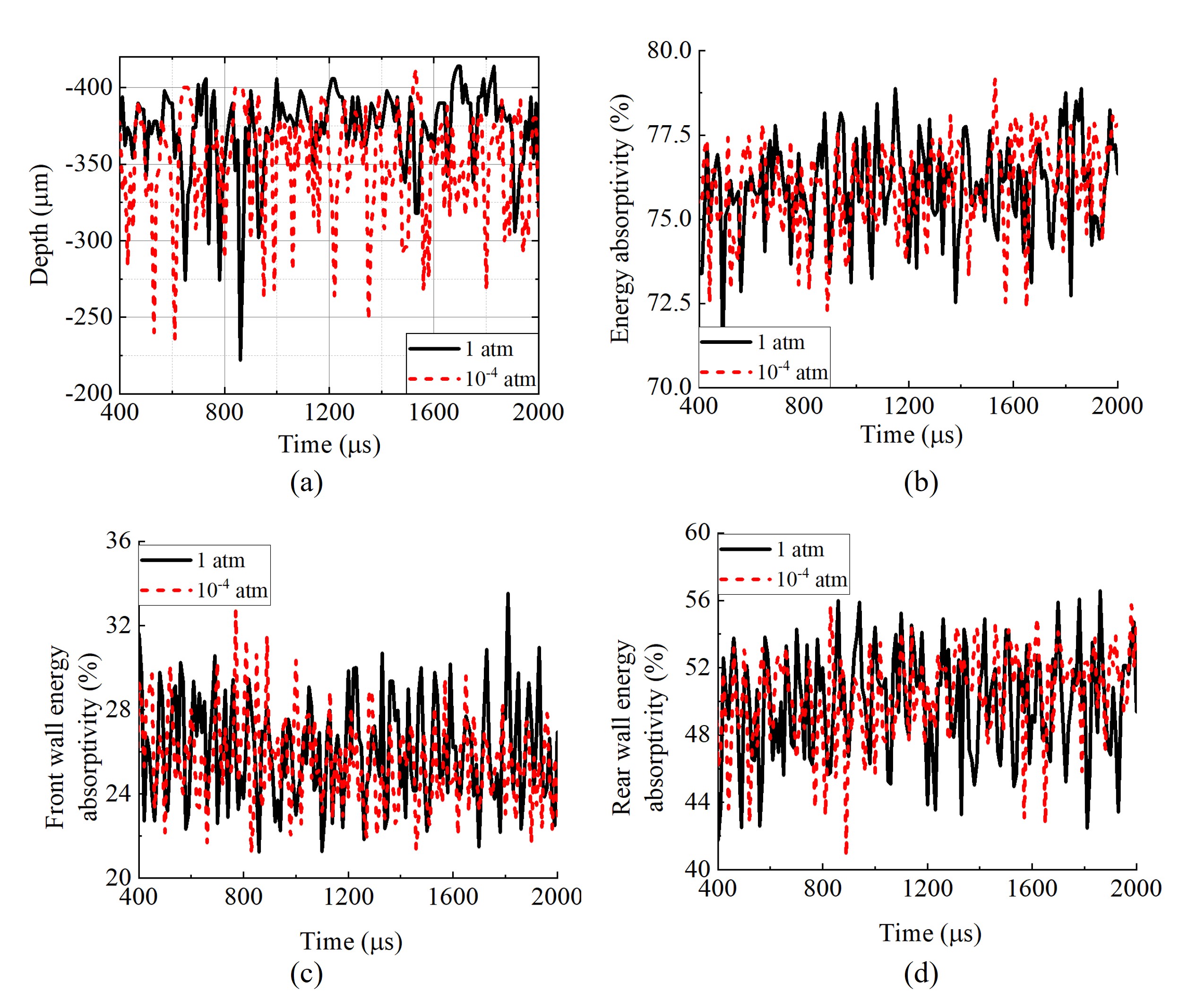}
	\caption{Comparison of Case3 ($1 \;\mathrm{atm}$)  and Case 5 ($10^{-4} \;\mathrm{atm}$): (a) keyhole depth, (b) total energy absorptivity, (c) energy absorptivity on the front keyhole wall, and (d) energy absorptivity on the rear keyhole wall.}
	\label{fig:abs_rec}
\end{figure}

To further analyze the stability of the keyhole under different ambient pressures, the keyhole depth and energy absorptivity of Case 3 and 5 are given in Fig.\;\ref{fig:abs_rec}.
The keyhole depth in Case 5 fluctuates more violently than that in Case 3.
The standard deviation of keyhole depth in Case 5 is $37\;\mathrm{\mu m}$, about 28\% higher than that in Case 3 as listed in Table\;\ref{tab:key_depth_pore}.
However, the keyhole depth is not indicative of the keyhole instability under the low ambient pressure. 
Further analysis of the energy absorptivity on keyhole surface from Fig.\;\ref{fig:abs_rec} (b-d) and Table\;\ref{tab:absorptivity} shows that the absorptivity fluctuations trend under different ambient pressures are similar to each other, whereas the standard deviations of the energy absorptivity on both the front and real keyhole wall in Case 5 are lower than those in Case 3 and even lower than those in Case 2.
As analyzed in Sect.\;\ref{S:3_3}, the keyhole stability is sensitive to the manufacturing parameters, which also explains why there is no keyhole pore in Case 5.
The energy absorption is calculated via the ray-tracing method, which is largely determined by the geometric shape of the keyhole.
Therefore, the lower standard deviation of the energy absorptivity is indicative of a smaller distortion of the keyhole shape, especially the rear keyhole wall.

Furthermore, the similar energy absorptivity on the rear keyhole wall under the low ambient leads to higher evaporation and higher recoil pressure than that under common ambient pressure.
The analysis above indicates that the keyhole depth fluctuation is more volatile under low ambient pressures because the evaporation is greater compared to that under common ambient pressure.
Simultaneously, the recoil pressure on the rear keyhole wall increases to maintain the stability of the keyhole shape.
Thus, the keyhole is stabler under low ambient pressure to reduce the porosity in the keyhole melting mode.

\section{Conclusions}
Keyhole pore formation during metal additive manufacturing has been investigated in this work with a multiphysics thermal-fluid flow model, and directly validated with the X-ray imaging results.
The following conclusions can be drawn:
\begin{enumerate}[label=(\arabic*)]
    \item The keyhole pore formation process has two distinct stages: (i) the instant bubble formation and (ii) pinning on the solidification front stage. 
    The instant bubble formation is mainly due to the keyhole instability (unbalanced forces on the rear keyhole wall).
    During the bubble pinning on the solidification front, the high flow speed below the instant bubble generates a vertical drag force which impedes the bubble from floating up to the molten pool surface.
    The bubble is finally caught by the solidification front to form the keyhole pore.
    \item The unevenly distributed recoil pressure on the keyhole surface increases the possibility of keyhole collapse to form keyhole pores.
    Additionally, the drag force from the mushy zone is pertinent as it determines the keyhole fluctuation at the bottom of the melt pool.
    A Darcy drag force model with the consideration of grain morphology in AM is required to improve the accuracy of drag force and keyhole fluctuation calculation.
    \item The keyhole pore size is sensitive to the manufacturing parameters.
    As the laser scanning speed slightly increases, the  keyhole pore sizes decrease significantly, and the shape of the keyhole pore becomes spherical and horizontally distributed at the molten pool bottom.  Moreover, the features of keyhole fluctuation and energy aborptivity variation could be criterions to predict the likelihood of keyhole pore formation.
    \item Low ambient pressure is a feasible way to reduce or even eliminate the formation of keyhole pores.
    Compared to the common ambient pressure, the recoil pressure on the rear keyhole wall under low ambient is larger and maintains a stable keyhole shape.
\end{enumerate}

\section*{Acknowledgments}
This research is supported by A*STAR under its AME IRG Grant (Project No. A20E5c0091).
\appendix
\section{Material Property}
\begin{table}[H]
	\small
	\centering
	\caption{Material properties of Ti-6Al-4V \cite{  wang2020evaporation,bayat2019keyhole,yan2017multi, klassen2014evaporation,tan2013investigation}} 
	\label{tab:thermal-CFD-values}
	\begin{tabular}{l l} 
		\hline
		Property &    Ti-6Al-4V  \\ 
		\hline   
		Solidus temperature ($T_s$) & 1878 [K] \\
		
		Liquidus temperature ($T_l$) & 1928 [K] \\
		
		Boiling temperature ($ T_b $) & 3315 [K] \\
		
		Solidus Density ($\rho$)  &  $ 4400 \rm{[kg/m^3]}$ \\
		
		Latent heat of melting ($L_m$) & $2.86\times 10^{5}$ [J/kg] \\
		
		Latent heat of evaporation ($L_v$) & 9.7$\times 10^{6}$ [J/kg] \\
		
		Saturated vapor pressure ($P_{e}$)   & 1.013$\times 10^{5}$ [Pa] ($T_b$=3315 K) \\
		
		Solidus Specific heat ($ c_s $)& 570 [J/(K$\cdot$kg)] \\
		
		Liquidus Specific heat ($ c_l $)& 831 [J/(K$\cdot$kg)] \\
		
		Thermal conductivity at solidus ($k_s$) & 16 [W/(m$\cdot$K)] \\
		
		Thermal conductivity at liquidus ($k_l$) & 32 [W/(m$\cdot$K)] \\
		
		Surface radiation coefficient ($ \epsilon $) & 0.4 \\
		
		Surface tension coefficient ($\sigma_0$)& 1.68 [N/m] \\
		
		Temperature sensitivity of $\sigma$ ($\sigma_s^T$)  & 0.00026 \\
		
		Viscosity ($\mu$) & 0.005 [Pa$\cdot$s] \\
		\hline
	\end{tabular}
\end{table}




 \bibliographystyle{model1-num-names}
\bibliography{elsarticle-template-1-num.bib}







\end{document}